\documentclass{article}

\usepackage{arxiv}

\usepackage[utf8]{inputenc} 
\usepackage[T1]{fontenc}    
\usepackage{hyperref}       
\usepackage{url}            
\usepackage{booktabs}       
\usepackage{amsfonts}       
\usepackage{nicefrac}       
\usepackage{microtype}      
\usepackage{lipsum}		
\usepackage{graphicx}
\usepackage{natbib}
\usepackage{doi}
\usepackage{placeins}
\usepackage{times}
\usepackage{epsfig}

\renewcommand{\shorttitle}{\textit{arXiv} Template}

\title{Can autism be diagnosed with AI? A narrative review}

\author{ Ahmad Chaddad $^{1,3}$*, Jiali li $^{1,}$, Qizong Lu $^{1}$, Yujie Li $^{1}$, Idowu Paul Okuwobi $^{1}$\\
\textbf{Camel Tanougast} $^{2}$,\textbf{Christian Desrosiers} $^{3}$ and \textbf{Tamim Niazi} $^{4}$ \\
$^{1}$\quad School of Artificial intelligence, Guilin Universiy of Electronic Technology, Guilin, Guangxi, China \\
$^{2}$ \quad University of Lorraine, Metz, Lorraine, France \\
$^{3}$ \quad The Laboratory for Imagery, Vision and Artificial Intelligence, Montreal, Quebec, Canada \\
$^{4}$ \quad Lady Davis Institute for Medical Research, McGill University, Montreal, Quebec, Canada \\
email:ahmadchaddad@guet.edu.cn}

\begin{document}
\maketitle

\abstract{Radiomics with deep learning models have become popular in computer-aided diagnosis and have outperformed human experts on many clinical tasks. Specifically, radiomic models based on artificial intelligence (AI) are using medical data (i.e., images, molecular data, clinical variables, etc.) for predicting clinical tasks like Autism Spectrum Disorder (ASD). In this review, we summarized and discussed the radiomic techniques used for ASD analysis. Currently, the limited radiomic work of ASD is related to variation of morphological features of brain thickness that is different from texture analysis. These techniques are based on imaging shape features that can be used with predictive models for predicting ASD. This review explores the progress of ASD-based radiomics with a brief description of ASD and the current non-invasive technique used to classify between ASD and Healthy Control (HC) subjects. With AI, new radiomic models using the deep learning techniques will be also described. To consider the texture analysis with deep CNNs, more investigations are suggested to be integrated with additional validation steps on various MRI sites.}


\section{Introduction}
Autism spectrum disorder (ASD) is a pervasive developmental disorder with cognitive abilities that are below normal for their age group. Its core symptoms are categorized by social communication deficits, repetitive stereotypical interests, and persistent patterns of behavior \cite{3}. For example, ASD patients have an inability to understand other intentions properly, reduced interactive eye contact, etc. Specifically, ASD endangers the physical and mental health of children, placing a burden on patients' social interaction, learning, life, employment, family, and society \cite{van2021age}. In this context, early diagnosis and early intervention for children with ASD can greatly improve the lives of those affected \cite{4}. Unfortunately, ASD has unclear direct indicators, and many studies have suggested the genetic factors \cite{1}, immunological \cite{17}, and neuropsychological associations \cite{18}. The prevalence of mental disabilities among children aged six and below in China is 1/1000, and ASD accounts for 36.9\% of them \cite{11}. According to the World Health Organization, the prevalence of ASD is gradually increasing worldwide, with a global average prevalence of 62/10,000 (0.62\%), equivalent to one child with ASD in every 160 children. It has become a major global public health problem \cite{12}.

Most of the diagnostic tools and methods are based on various tests. The present research work on ASD symptoms is not accurate \cite{kim2018variability}. It is generally believed that children with ASD will have many problems in their growth and development \cite{2019Parenting,2015Diagnostic}. So far, there are incomplete diagnostic tools for ASD. In this context, a multi assessment is usually considered. ASD could be related to developmental delays and  abnormalities. This assessment considers the growth history, parental interview \cite{2019Parenting}, medical examinations if necessary, and many other things. According to \emph{American Psychiatric Association's Diagnostic and Statistical Manual of Mental Disorders, 5th edition (DSM-5)}, the diagnostic criteria for ASD consist of persistent defects in social communication, social interaction, restricted and repetitive behavior patterns \cite{2015Diagnostic}. In \cite{2000The}, the Autism Diagnostic Observation Schedule (ADOS) is considered the gold standard test because of its reliability, validity and usefulness. Unfortunately, this test-based screening method can only diagnose children when they have the ability to communicate. We note that there is a great advantage when ASD could be identified at an earlier age. However, according to traditional methods, ASD is difficult to identify at an early age \cite{10} due to the gaps in cognitive abilities in infants at 24 months or older \cite{9}. Even with clinical investigation, deep neurological assessment seems to be more needed. ASD diagnosis is improved by involving many neurological techniques/features (e.g., brain waves  \cite{6,heunis2018recurrence}, magnetic resonance images (MRI) \cite{8} and eye-tracking techniques \cite{7}, etc.). Another aspect of assessment is related to many genes for predicting ASD \cite{cederquist2020multiplex}. For example, SHANK3 \cite{2013Prospective} and PTCHD1 \cite{2010Disruption} are two genes involved in the pathogenesis of ASD through regulation of the nervous system. ITGB3 \cite{2011Family}, is associated with the pathogenesis of similar disorders. Like genomic analysis, imaging analysis is a promising technique that leads to identify ASD patients. 

Imaging like MRI is used to show the anatomy of brain (e.g., ASD patients) \cite{23}. Two types are the most considered identifying ASD patients: 1) functional magnetic resonance imaging (fMRI) and 2) structural magnetic resonance imaging (sMRI or MRI) \cite{sen2018general}. The fMRI can show brain function, like active brain regions. While sMRI shows the structure variations (e.g., growth, deformation, atrophy, etc.). In addition, sMRI/MRI is currently the most used technique for imaging the brain structure due to its fast and high-resolution 3D volume imaging. For ASD, sMRI can describe structural brain changes by analyzing gray matter volume, cortical thickness, cortical complexity, and co-variance networks. While fMRI relies on the oxygen content of local tissue vessels (blood oxygen levels depend on functional brain MRI imaging) and can track signal changes in real time. So far, MRI or fMRI provide relevant imaging features that are related to ASD \cite {sen2018general}.

Imaging features (or radiomics) are widely used in medical image analysis. Among the imaging feature techniques, the following features (radiomics) are the most features used for ASD, namely, 1) colour features, 2) texture, 3) shape/morphology, and 4) spatial relationship features. Briefly, $colour features$ describe the surface properties of the image. It is not affected by image rotation and translation; $texture features$ can better describe the structure image; $shape features$ can effectively describe the geometrical area (i.e., region of interest); $spatial relationship features$ can enhance the ability to describe and distinguish the content of the image. These features have been used for many clinical applications such as cancer \cite{chaddad2019radiomics,chaddad2018novel,chaddad2018radiomics,7591612, chaddad2014brain, chaddad2019integration, chaddad2019predicting,chaddad2017predicting}, neuroimaging \cite{zhang2021autoencoder,chaddad2018deep}, segmentation \cite{chaddad2016quantitative}, etc. As ASD examples, MRI regional features were computed to study the abnormalities in brain development of ASD patients \cite{YangLiu}. In \cite{Xiuyan}, shape features are considered to predict ASD. In \cite{2016Cortical,wolff2018journey}, multiple ASD brain developmental abnormalities are detected during infancy. Moreover, many studies have shown that young children with ASD have a much larger brain size compared to their normally developing peers \cite{1997Macrocephaly}. The overall volume and density are significantly larger than those of normal children (or healthy control) \cite{2015Head}. Briefly, we will describe the brain differences between ASD and HC as follows:

\emph{Surface area}: Studies have shown that the early cerebral cortex of children with autism expands rapidly between six and twelve months. This atypical expansion leads to problems such as visual receptivity deficits and neglect of social cues \cite{2013White}. Cortical surface area increases with an accelerated rate between one and two years of age \cite{13}. It coincides with problems of social deficits. Another longitudinal study found that the white matter of the temporal lobe of the brain increased in autistic children between two and four to five years old. However, the brain grows at a similar rate to normal during this age interval \cite{14}. As a result, the overgrowth of the temporal and frontal lobes is an indicator of ASD \cite{12,13,14}.

\emph{Cerebrospinal fluid}: An excessive increase in cerebrospinal fluid can also occur in young children with ASD. During the first few months of life (i.e., within 24 months), infants with ASD have a high volume of inter-axial extra-cerebrospinal fluid \cite{999}. Specifically, the increase in the volume of extra-axial cerebrospinal fluid is more significant at 6 months, which is about 25\% higher than ordinary babies. This is related to movement, communication, and the condition of ASD. When the extra-axial cerebrospinal fluid continues to rise, the communication disorder will become more serious \cite{998}.

\emph{Structural abnormalities in the white matter}: The corpus callosum develops abnormally at six months. Its area and volume increase significantly \cite{15}. This abnormality is positively correlated with the stereotypical behavior of children with autism \cite{16}. Therefore, structural abnormalities in the white matter are likely to be an important causal factor in the core social deficits (especially emotional disturbances) of later autism. 

For ASD diagnosis, clinicians have been committed to using neuroimaging tools. It can automatically distinguish patients with brain diseases from HC or other patients. This can be achieved using features (e.g., imaging, genes, clinical, omics, etc.) with machine learning (ML). ML consists of many methods to classify between classes (e.g., neural networks, support vector machines, random forests, etc.). It learns how to identify the features associated with ASD and then constructs a relevant model. The accuracy of a classifier/predictive model is improved by training the model on large datasets. Eventually, the model can be relied upon to diagnose the presence of ASD. Its accuracy is measured by how well it is able to predict the true class (e.g., ASD). In this context, we aim in this paper to discuss the general radiomics/features and AI/ML model for predicting ASD.

The rest of this paper is structured as the following: Section 2 contains a review of the literature on ASD diagnostic methods. We then discuss the general radiomic methodology for predicting ASD in Section 3. In Section 4, we present the recent explainable artificial intelligence (XAI) literatures that are related to ASD. Section 5 discusses the strengths and limitations of ASD predictive models and summarizes the main findings of this study. Last, Section 6 concludes the paper with future recommendations.

\section{Related works}
Increasing attention has been remarkable for ASD, when Leo Kanner has talked about ASD in 1943, and has mentioned that the ASD is related to the brain \cite{26,27,29,30}. In \cite{26}, children with ASD show larger brain volume than HC. In \cite{27}, ASD was related to large and small brain white matter hyperplasia and early gray matter hyperplasia, respectively. Compared to HC, ASD children have a larger volume in the amygdala \cite{29} and hippocampus \cite{30, 31}. Most of these studies consider the classifications between ASD and HC according to the difference in brain volume or thickness. While, texture feature based on gray-scale co-occurrence matrix (GLCM) and Laplacian filter was firstly appeared by Chaddad et al. to compare \cite{8} and classify \cite{32} between ASD and HC. As the prevalence of autism increases year by year, effective ASD diagnostic methods have become a major concern worldwide. We summarize three main diagnostic methods as follows:   

\emph{Electroencephalography (EEG)}: EEG measures neural activity and can detect children at risk of developing ASD and, thus, provide an opportunity for early diagnosis. For example, EEG data is used to compare between ASD and HC \cite{19,heunis2018recurrence,vicnesh2020autism}. In \cite{20}, used CNN model for classification after converting the data into 2D form. Although EEG can be used to diagnose ASD, it still has limitations in a number of conditions (e.g., signal noises). 

 \emph{Eye tracking}: It is based on characteristic changes of eyes, like periphery and iris. In \cite{21}, they studied 86 two-year-olds (26 ASD, 38 HC and 22 Developmentally delayed children). It shows that eyes with ASD were associated with passive insensitivity to social signals. In \cite{22}, they selected 29 ASD children aged between 5-11 years. Through the visualization of real faces and avatars, it was possible to study how children with ASD recognize emotions. For ASD patients, eye tracking is not an optimal method because it takes long time for children to cooperate, in addition that it is not flexible for clinical diagnosis.
 
 \emph{MRI/fMRI scans}: Data quality of MR imaging is improving in function with the advanced technology. Previous studies have shown that brain structures in patients with ASD can differ significantly in terms of volume, thickness, and texture \cite{25,32}. However, this scenario still under radiomic and AI investigation for ASD diagnosis. Yet, no clear tools have been involved in the clinical system. However, we find many works prove that classifier models using extracted features from MRI/fMRI images have the ability to predict ASD patients. For example, a support vector machine (SVM) model has shown an accuracy value of 66.8\% to predict ASD images \cite{2018AAA}. In \cite{8759193}, 12 classifiers are compared, namely, six nonlinear shallow ML models, three linear shallow models, and three deep learning models. A dense feedforward network provides the best results among the 12 models with AUC value of 80\%. This demonstrates that even when using features derived from imaging data, deep learning methods like dense feedforward network can provide higher predictive accuracy over classical ML methods \cite{8759193}. To let the AI models be feasible for ASD prediction, more investigation is recommended since the performance metrics are still limited. In addition, no clear study is considered when the images from MRI scanner consist of 7 Tesla or above. We believe that more resolution of images will let the radiomic with texture analysis be more informative for predicting ASD \cite{8,32}.
 
\section{Radiomic methodology}
\begin{figure*}[htp]
    \centering
    \includegraphics[width=18cm]{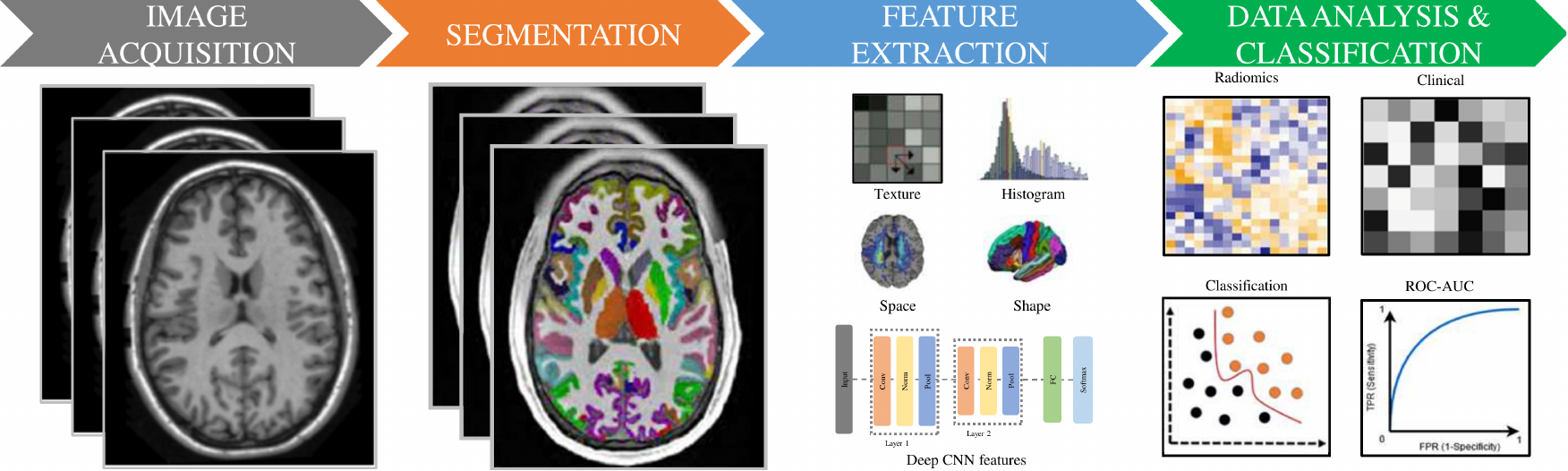}
    \caption{General radiomic workflow for predicting ASD. Schematic illustration of the entire radiomic process, including image acquisition with preprocessing with symptomatic ASD patients undergo MR (MRI and/or fMRI) scans. After image segmentation, radiomic features are extracted and selected. Data aggregation for statistical modelling, with classifier modeling employed for classifying ASD from HC.}
    \label{fig1}
\end{figure*}

To provide a wider perspective to the readers, radiomic pipeline is simply given in Figure \ref{fig1}. It illustrates the processing steps for radiomic pipeline that consist of image acquisition and preprocessing, segmentation, feature extraction, statistical analysis, and classifications. We described below a detailed review of each step.

\subsection{Image acquisition and preprocessing}
 The purpose of preprocessing is to improve the visual effect of the image. It can purposely emphasize the overall or partial characteristics of the image for various scenarios \cite{wang2020experiment}. For example, improve the color, brightness, and contrast of the image. There are two main methods for image enhancement: a) Spatial domain that includes image gray-scale transformation, histogram correction \cite{39}, local statistics method, image smoothing and image sharpening, etc. As in \cite{37}, fast non-rigid registration can improve the contrast of the brain structures. In \cite{38}, image enhancement method based on brightness level and gradient modulation. This method reduces the dynamic range of the brightness level and enhances the details (i.e., texture) of the image. b) frequency domain that transforms an image into the frequency domain for filtering using Fourier analysis. The image is then inversely transformed back into the spatial domain. The most used frequency-domain methods are the homomorphic filtering \cite{yugander2020mr} and wavelet transform \cite{sagheer2020review}. Therefore, image denoising plays a major role for texture analysis. It can be described by probability distribution function and probability density function. We note that the texture analysis could be related to the scale of image filtering and then to ASD  \cite{park2017high}. 
 
\subsection{Normalization/standardization}
Image normalization is one of the preprocessing steps to avoid image distortions (i.e., translation, rotation, scaling, and skew). However, the main challenge for ASD studies is related to MRI images derived from multisite \cite{yankowitz2020evidence, hoeksma2005variability, delisle2021realistic}. For example, scans from multisite lead to high differences of texture features between these sites. In \cite{121}, Min-Max normalization is used to overcome the image variation, and convert the image values to a range of [0-1]. Normalization of images here improves the learning rate, reduces the dependence on initialization, and reduces the training time to overcome the overfitting problem. In \cite{125}, non-parametric models are used to correct intensity inhomogeneities and avoid the scanner distortions. In \cite{122}, MRI was normalized using voxel-based morphometry (VBM) that is available in the Statistical Parametric Mapping software \footnote{$https://www.fil.ion.ucl.ac.uk/spm/software/$}. VBM technique is spatially normalizing the MRI scans to the same stereotactic space to correct nonuniform intensity variations \cite{ashburner2000voxel}. Recent work shows that domain adaptation can effectively reduce the site variation using the CNN models \cite{delisle2021realistic}. However, more work on image normalization for multisite variation is needed to extract the texture features for radiomic analysis.

\begin{table*}[ht!]
\caption{Summary of ASD studies using MRI/fMRI with machine learning models.}\label{table1}
\renewcommand{\arraystretch}{1.2}
\tiny
\begin{tabular}{lccccccccc}
 \toprule
\textbf{Work} & \textbf{Data Source} & \textbf{Cases Number} & \textbf{Data Type} & \textbf{FEM}& \textbf{Classifer Model}& \textbf{Acc} & \textbf{Sen}& \textbf{Spec} & \textbf{AUC}   \cr
\hline
\cite{46} & FSL & 50 ASD and 50 HC & TfMRI & SPF & DWT-CNN & 80\% & 84\%  & 76\% & - \\
\cite{47} & ABIDE-I+II &  23 ASD and 15 HC & Rs-fMRI & SPF & SVM & 80.76\% & -  & - & - \\
\cite{48} & NDAR & 185 subjects &  sMRI-fMRI & SF & RF & 80.8\% & 84.9\%  & 79.2\% & 81.92\% \\
\cite{49} & ABIDE & 505 ASD and 530 HC &   Rs-fMRI & SPF & Ridge Return & 71.98\% & -  & - & - \\
\cite{50} & ABIDE & 518 ASD and 567 HC & rs-fMRI & SF &  CNN  & 71.8\% & 81.25\%  & 68.75\% & 67\% \\
\cite{51} & Private & 40 ASD and 36 HC &  MRI & SF & SVM   & 84.2\% & 80\%  & 88.9\% & - \\
\cite{52} & ABIDE-I & 505 ASD and 530 HC &  rs-fMRI & OF & DNN    & 70\% & 74\%  & 63\% & - \\
\cite{53} & ADHD-200 & 279 ASD and 279 HC  & fMRI & TF & SVM   & 64.91\% & 44.16\%  & 81.91\% & - \\
\cite{55} & ABIDE-I+II & 76 ASD and 75 HC & MRI & TF and SF & SVM   & 64.3\% & 77\% & 82\% & 69\% \\
\cite{56} & ABIDE-I & 155 ASD and 186 HC & T1-MRI & SF &  HGNN  & 76.7\% & -  & - & - \\
\cite{57} & ABIDE-I+II & 255 ASD and 276 HC & rs-fMRI & SF &  SVM  & 75.00\%–95.23\% & 90.62\%  & 90.58\% & - \\
\cite{58} & ABIDE & 539 ASD and 573 HC & T1-MRI & SF &  6 classifiers  &  $>$80\% & -  & - & - \\
\cite{90} & ABIDE & 539 ASD and 573 HC & rs-fMRI & OF & SVM  &  86.7\% & 87.5\% & 85.7\% & - \\
\cite{91} & ABIDE & 99 ASD and 85 HC & fMRI & SPF & CNN  &  68.54\% & 69.49\% & 67.58\% & - \\
\cite{92} & ABIDE-I & 270 ASD and 305 HC & rs-fMRI & SPF & ANN & 74.54\% & 63.46\% & 84.33\% & - \\
\cite{93} & ABIDE-I & 48 ASD and 24HC  & MRI & TF and SF & RF   & 98\% & - & -  & 52.5\%- 53\% \\
\cite{94} & ABIDE & 49 ASD and 41 HC & rs-fMRI & SF &  SVM  & 78.89\% & 85.71\% & 70.73\% & - \\
\cite{95} & ABIDE & 539 ASD and 573 HC & fMRI & SF &  CNN  & 87\% & - & -  & - \\
\cite{69}& ABIDE-I & 505 ASD and 530 HC  & fMRI & SF & CNN & 70.22\% & 77.46\% & 61.82\% & 74.86\% \\
\cite{68}& ABIDE-I & 79 ASD and 105 HC  & 3D-fMRI & OF & CNN & 94.7\% & - & - & 94.703\% \\
\cite{57}& ABIDE-I+II & 255 ASD and 276 HC  & rs-fMRI & SF & SVM-RFECV & 75.0\%-95.23\% & 90.62\% & 90.58\% & -\\
\cite{72}& ABIDE-I & 368 ASD and 449 HC  & sMRI & SF & AE, MLP & 85.06\% & - & - & - \\
\cite{71}& ABIDE-I+II & 620 ASD and 542 HC  & rs-fMRI & SF &  3D-CNN,SVM & 72.3\% & - & - &-  \\
\cite{75}& ABIDE-I & 505 ASD and 530 HC & rs-fMRI & OF & CNN & 82.69\% & 88.23\% & 88.67\% &-  \\
\cite{73}& ABIDE-I & 403 ASD and 468 HC  & fMRI & OF & SVM & 76.8\% & 72.5\% & 79.9\% & 81\%  \\
\cite{79}& ABIDE-II & 26 ASD and 26 HC  & MRI & SF & SVM-RFE & 73\% & 71\% & 75\% & 81\% \\
\cite{78}& ABIDE-I & 403 ASD and 468 HC  & rs-fMRI & SF & RNN-LSTM & 74.74\% & 72.95\% & - &-  \\
\cite{77}& ABIDE-I & 505 ASD and 530 HC  & fMRI & SF & SAE & 70.8\% & 62.2\% & 79.1\% &-  \\
\cite{76}& ABIDE-I & 505 ASD and 530 HC  & sMRI & SF & RFE+RF & 72\% & - & - &-  \\
\cite{86}& ABIDE-I & 368 ASD and 449 HC  & sMRI & SF & AE & $85.06\pm3.52\%$ & - & - &-  \\
\cite{85}& NDAR & 47 ASD and 24 HC & rs-fMRI  & OF & SVM-RFE & 86\% & 81\% & 88\% &-  \\
\cite{84}& ABIDE & 539 ASD and 573 HC  & fMRI & SF & RCE-SVM & 70.01\% & - & - &-  \\
\cite{83}& ABIDE-I & 539 ASD and573 HC  & rs-fMRI & SF & SVM & 86.7\% & 87.5\% & 85.7\% &-  \\
\cite{82}& NDAR & 33 ASD and 33 HC  & fMRI & SF & 1D-CNN & 77.2\% & 78.1\% & 76.5\% &-  \\
\cite{81}& ABIDE & 41 ASD and 41 HC  & rs-fMRI & OF & KNN & 85.9\% & 79.3\% & 92.6\% &-  \\
\bottomrule
\end{tabular}
\tiny\\
\label{tab:multicol}
SPF: spatial feature, TF: texture features, SF: shape/morphological feature, OF: other features, rs-fMRI: resting state-functional magnetic resonance imaging, T1-MRI: T1-weighted magnetic resonance imaging, 
ABIDE I and II: autism brain imaging data exchange I and II, NDAR: national database for autism research, FSL: fMRI software library, SVM-RFECV: support vector machine-recursive feature elimination with a stratified-4-fold cross-validation, AE: autoencoder, SAE: stacked autoencoder, DWT-CNN: discrete wavelet transform-convolutional neural network, MLP:muti-layer perceptron, DNN: deep neural networks, KNN: kohonen neural network, RNN: recurrent neural networks, LSTM: long short-term memory, SVM-RFE: support vector machines-recursive feature elimination, RCE-SVM: recursive cluster elimination-support vector machines, FEM: feature extraction method, Acc: accuracy, Sen: sensitivity, Spec: specificity, AUC: area under curve, HGNN: hypergraph neural network, +: combination, >: is greater than, ±: plus or minus.
\end{table*}

\subsection{Segmentation/labeling}
ASD Segmentation aims to label the brain regions (e.g., region of interest). An image is divided into several regions with similar properties. Currently, clustering and deep learning techniques are the main methods for segmenting brain MRI images \cite{42}. For example, deep learning is able to segment properly the corpus callosum (CC) \cite{108}. This technique reduces the need for manual or semi-automatic segmentation of neuroanatomical. Manual and semiautomatic segmentation can be performed on brain MRI using 3D Slicer tool \cite{pieper20043d}. However, deep learning based segmentation offered significant algorithms for labeling brain regions in automatic fashion \cite{dolz20183d}. 
Actually, the most used tool for ASD image preprocessing, standardization and brain region labeling is the Freesurfer \cite{fischl2012freesurfer}. Specifically, it processes the 3D structure brain image, performs automatic cortical and subcortical labeling. By generating accurate gray and white matter, and cerebrospinal fluid regions, it can compute cortical thickness and other surface characteristics. Specifically for ASD, FreeSurfer is widely used in the preprocessing of MRI images. For example, it is used to preprocess and extract features from MRI images of ASD patients \cite{109}. While it is analyzed, high-quality MRI images in \cite{111}. It is also considered generating brain morphological features, including regional volume, surface area, average cortical thickness and Gaussian curvature  \cite{112}.

\subsection{Features Extraction}
Image consists of many features that define the behavior of an image \cite{6783417}. Specifically, feature extraction techniques aim to find the most important information to save computational work and data storage. Briefly, we summarize three types of image features that used for predicting ASD: 1) shape features, 2) spatial features, and 3) texture features.

\emph{Shape features}: This type of features is related to the geometric and morphological region of interest (e.g., brain subcortical regions). For example, many studies consider the shape features to predict ASD patients \cite{27,29,30}. The shape feature problem is represented by unreliable results when the target is deformed, in addition to distortion due to changes in viewpoint. We note that Hough transform and Fourier shape descriptors are classical methods to extract shape features. Despite the wide use of shape features, this type of features do not describe the content of the image. Then, in combination with other informative features, may improve performance metrics \cite{8,32}

\emph{Spatial features}: It refers to the spatial position or relative direction. It can strengthen the description and distinction of image content. While, the rotation and change of the scale can affect the spatial characteristics. There are two ways to extract the spatial relationship: 1) extract the features using the automatic segmentation (objects, colors, etc.) and 2) by generating an index. Alternatively, you can segment the image uniformly, extract the features from each image separately, and consider the index. What's more, spatial features have advantages in diagnosing ASD patients of different ages and genders. For example, spatial filter can provide highly discriminative features between ASD patients and neurotypical subjects \cite{128}. However, there are few studies that used spatial features for ASD diagnosis due to the constraint of high-dimensional data and a relatively small data set \cite{129}. 

\emph{Texture feature}: Most of the current literatures for predicting ASD based on images  are based on shape features. However, the potential of MRI images has not been fully developed. Fortunately, studies like \cite{8,32} have proved that texture features can classify patients with ASD. Specifically, texture features reflect the homogeneity of the image \cite{8,32}. These features describe the surface properties of the image \cite{45}. Specifically, the texture is based on a statistical order that is widely used for many topics. For example, gray-level co-occurrence matrix (GLCM) is currently one of the best statistical techniques for computing image texture. Computation of GLCM reflects comprehensive information about the direction, adjacent interval, and gray level of an image. In addition, the local patterns and their arrangement rules are analyzed using this technique.

Table \ref{table1} reports the performance metrics in predicting the ASD  using the current feature extraction techniques and various image sources (ABIDE \footnote{$http://fcon_1000.projects.nitrc.org$}, NDAR \footnote{$https://nda.nih.gov$} and FSL \footnote{$https://www.fmrib.ox.ac.uk/datasets/$}). We found that shape/morphological  and texture features lead  generally to higher accuracy rate comparing to shape or texture features. We noted that the use of texture features is still limited due to the limited resolution of MRI/fMRI scans with 1.5 or 3 Tesla. Thus, more investigation related to texture analysis is recommended for improving the performance metrics.

\begin{table*}[ht!]
\centering
\tiny
\caption{Summary of feature selection techniques related to ASD.}\label{table2}
\setlength{\tabcolsep}{6pt}
\renewcommand{\arraystretch}{1.2}
\begin{tabular}{lccp{9cm}}
 \toprule
    \multicolumn{1}{l}{\textbf{Work}} & 
    \multicolumn{1}{c}{\textbf{Feature group}} & 
    \multicolumn{1}{c}{\textbf{Feature selection type}} & 
    \multicolumn{1}{l}{\textbf{Technique}}  \cr
\hline
\cite{51} & SF & WM & Identify the feature group that achieves the best performance through greedy forward feature selection.\\
\cite{136} & OF & WM & A feature selection algorithm based on a minimum spanning tree is proposed to find the optimal feature set.\\
\cite{137} & SF & WM & Use recursion to perform feature selection.\\
\cite{138} & OF & FM & Use Pearson correlation coefficient to filter redundant features.\\
\cite{139} & OF & WM & Use recursive feature elimination (RFE) to rank the importance of features and then remove irrelevant features recursively.\\
\cite{140} & OF & WM & Use the reverse order feature selection algorithm.\\
\cite{141} & OF & WM & Adopt a restricted path depth-first search algorithm (RP-DFS).\\
\cite{142} & OF & FM & Chi-Square is used to remove non-significant features.\\
\cite{143} & SF & EM & Use principal component analysis (PCA) to select the principal components.\\
\cite{144} & SF & EM & Use the sure independence screening (SIS) method. Multiple features are removed in each iteration.\\
\bottomrule
\end{tabular}
\tiny\\
\label{tab:multicol}
SF: shape/morphological feature, OF: other features, FM: Filter method, WM: Wrapper method, EM: Embedded method
\end{table*}

\subsection{Feature selection}
Due to the high-dimensional nature of MRI data, features may consist of redundant information \cite{136}. Feature selection is a procedure to choose the dominant features. Specifically, feature selection algorithms aim to find the most predictive features by removing irrelevant or redundant features. This procedure improves the classifier model performance and reduces the running time \cite{133}. These algorithms can be classified to three methods, namely, filter, wrapper, and embedded \cite{130,132}.

\emph{Filter method (FM)}: The features of each dimension are given weights, which represent the importance of the feature.  These features are then ranked according to the weights \cite{131}. A number of features are selected using a threshold. The typical methods are Pearson correlation coefficient and chi-square test. This kind of feature selection algorithm has low algorithmic complexity and is suitable for large-scale data sets. However, it has a lower classification performance compared to wrapper algorithms.

\emph{Wrapper method (WM)}: It divides the features into different combinations, evaluates the combinations, and compares them with other combinations. Typical methods are represented by recursive feature elimination (REF), step-wise selection, backward elimination, etc. These algorithms are convenient with some studies. Despite the advantage of wrapper methods, more investigations to generalize these algorithms are needed \cite{134}.

\emph{Embedded method (EM)}: The feature selection algorithm itself is embedded in the learning algorithm as a component \cite{130}. ML models are used for training, then obtain the weight coefficients of each feature. Features are selected based on coefficients from the largest to smallest (similar to the filter method, except that the coefficients are trained). This method is considered as an efficient technique to select predictive features. 

Table \ref{table2} reports the techniques used recently for feature selection in ASD studies. For example, wrapper methods  are generally more used comparing to filtering and embedding. As expected that the predictive feature derived WM is higher performance than FM \cite{146}. Due to the difficulty of setting parameters, the use of EM is limited. More details about feature selection techniques are reported in \cite{145}.

\subsection{Statistical analysis and classification models}
To predict the ASD images, features extracted are used as input to a classifier model. Many ML models could be used as predictive models. The ML models are generally divided into two types: supervised and unsupervised \cite{nielsen2020machine}. An algorithm based on supervised learning uses labeled input and output data, while an unsupervised learning algorithm does not. We summarize two groups of ML methods: conventional methods (i.e., SVM, KNN, RF, etc.) and deep learning (e.g., CNN, RNN, LSTM, etc.).

\emph{Conventional method}: The most popular is the SVM. It is in many neuroimaging tasks \cite{115, 51, 116}. However, SVM is not recommended when the samples less than the features number due to the overfitting. In this context, random forest can solve this problem by selecting automatically the features to build the classifier model \cite{grossard2020children}. RF combines random feature selection and bootstrap aggregation to build a collection of decision trees that exhibit controlled variation \cite{amit1997shape}. To tune the parameters of conventional models, a grid search on a validation set is considered. In addition, many types of validation steps like cross-validation are used to test the classifier model \cite{berrar2019cross}.

\emph{Deep learning}: Deep learning is a part of ML models, advanced with new hardware technology like graphics processing unit (GPU). Recently, deep learning demonstrates remarkable classification results in clinical applications \cite{choi2018radiomics}. In \cite{64}, a hybrid model consists of CNN with brain features are used to improve the performance metrics. In addition, some literature combined the conventional methods and deep learning to improve the performance and overcome overfitting \cite{ thomas2020classifying}.

\emph{Statistical and performance metrics}: For evaluating the classifier models, many performance metrics are considered. However, the common measurements are the area under the receiver operating characteristic (ROC) curves (AUC), accuracy, sensitivity, and specificity. To compare between classes (e.g., ASD versus HC), significance tests are  used to measure the $p value$. Correction of significant values (e.g., $p < 0.05$) is recommended following the Holm-Bonferroni correction (or using other correction techniques) \cite{aickin1996adjusting}. For example,  we note that the range value of classification accuracy is 70.01-94.7\% depends on the features extracted, classifier model and data source (i.e., see Table \ref{table1}). We observed that the most common models are CNNs and SVMs. However, CNN demonstrates higher performance metrics comparing to other models \cite{114}. Moreover, ML or deep learning algorithms require a large data set to generalize a reliable predictive model, which is not available currently in a medical field. To get benefit from deep learning, a transfer learning technique is used to overcome the overfitting and time computation \cite{dominic2021transfer}. Although the potential of deep learning for clinical tasks, more work is required to understand the mechanism of such algorithms (e.g., information flow of CNN for classifications \cite{goldfeld2020information}).

\section{Explainable Artificial Intelligence}
Recent literatures for reporting clinical research that involves deep learning will realize the full potential of machine learning tools \cite{mateen2020improving,chen2021recent}. Unfortunately, these models (i.e., algorithms) work as a black box in the medical field \cite{105}. It is not explained how to correlate inputs and outputs or the mechanism of information flow in the hidden layers \cite{103}. XAI provides interpretability for algorithms, models, and tools. It aims to make AI algorithms more transparent to improve human understanding of these models. For example, CNNs can automatically extract features based on their convolutional layers, its interpretability is crucial for personalized diagnosis (e.g., ASD \cite{ruan2021deep}, Coronavirus \cite{chaddad2021deep}, etc.). The output can be mapped back to the input space to see which parts of the input are discriminative \cite{96}. In \cite{97}, loss function is considered for each filter within the high-level convolutional layer to produce interpretable activation patterns. In \cite{8767919}, convolutional layers of CNN models are quantified to understand the information flow from input to output of architecture for predicting Alzheimer disease using MRI images. EEG data were used to detect emotions in ASD patients, and an interpretable deep learning technique (SincNet) was investigated \cite{mayor2021interpretable}. In addition, an explainable SVM model for ASD identification was studied by demonstrating a link between the dominant features and the model outcome \cite{biswas2021xai}. 

Applying such XAI models in predicting ASD image will provide more details about the brain subcortical regions related to ASD. Most of the XAI is focused on model-agnostic post-hoc explainability algorithms due to their easier integration and wider reach \cite{lundberg2017unified}. Interpretable AI techniques can be generally characterized from a different perspective \cite{88}. While the former strategies are easier to grasp and hence adopt, their effectiveness is often limited, necessitating the deployment of more sophisticated procedures. Deep radiomic analysis, in which the CNN layers are encoded and utilized as input into a classifier model, is one of the most active study areas in XAI \cite{chaddad2021deep,8767919,chaddad2018deep}. In this context, deep radiomic analysis seeks to provide high-level transparency of deep learning algorithms in the health data (e.g., images). Despite gaining traction of XAI, evaluating these methods is still a challenge and poses an open question in the future of XAI research in clinical tasks.

\section{Discussions}
Using radiomics with AI models is considered a pioneering development of precision medicine work \cite{8767919,156} like in mental disorders (e.g., ASD) \cite{152}. This is motivating to make a systematic overview of the radiomic application for ASD diagnosis. There are only two classes (ASD and HC) available in the public domain, which are not able to investigate all subtypes of ASD. Although fMRI and sMRI data are public available in the ABIDE dataset, the results of combining these multisite data for ASD diagnosis using radiomics and deep learning models have not yet been investigated. As we previously mentioned that the texture features depend on the MRI sites that lead to bias when we combined all ABIDE sites. Nowadays, assistive tools using domain adaptation algorithms can reduce this issue; however, the problems still dominate when implementing these algorithms in real-world scenarios.

This study demonstrated the various uses of radiomic models in diagnosing and classifying ASD, along with their strengths and limitations. Critical examples of radiomic pipelines for ASD with classification accuracy, different evaluation measures, and essential feature selection and their techniques and dataset sources have been discussed and analyzed. However, certain problems prevailing need to be addressed, such as learning from limited data, considering inappropriate sampling methods, classification between imbalanced datasets and how we involve the XAI in radiomic analysis. Integrating AI in clinical settings would not only improve our knowledge of ASD, but will also allow healthcare practitioners to employ these methods as clinical decision support systems for screening and diagnostic processes.
To sum up, we summarize the main findings of this study on ASD as follows:
\begin{itemize}
    \item MRI-based models for the diagnosis of ASD are more suitable for clinical trials than eye-tracking and CT image analysis. MRI can provide more detail of the brain.

    \item The brain of ASD patients can be heterogeneous in many locations (e.g., hippocampus, amygdala, etc.). The variation could be captured by shape features (e.g., volume, thickness, etc.).

    \item Deep learning is still challenging to diagnose ASD patients due to the lack of benchmark datasets \cite{153}.
    
    \item XAI could be the solution as a diagnostic model for ASD. However, it needs more investigation in real-world scenarios.

    \item The public data set needs to be continually expanded to avoid inappropriate studies due to insufficient data. In addition, ensure that there is no error in results due to age, gender, etc.\cite{155}.
\end{itemize}

\section{Conclusions}
In this paper, we present a survey of AI related to ASD using MRI/fMRI scans. We discussed the general radiomic features and classifier models that are used for predicting the ASD images. Recent studies show that the texture features are informative features. Among the deep learning models, CNN demonstrates the highest metrics. However, more investigation is needed in the context of XAI. For future work, high-precision and high-transparency models can be established by quantifying the deep texture from CNN models to predict early ASD patients.

\bibliographystyle{unsrtnat}
\bibliography{references}
\end{quote}

\end{document}